\documentclass[11pt]{article}
\usepackage{latexsym} \usepackage{amsfonts}\usepackage{epsfig}
\begin{document}
\begin{center} {\large \bf  RENORMALIZATION PROPERTIES OF SOFTLY BROKEN
 \\[0.2CM]  SUSY GAUGE THEORIES\footnote{Talk presented at the conference
"Continuous Advances in QCD 2002/Arkadyfest", Minnesota, May
2002.} } \vspace{1cm}

{\large \bf Dmitri KAZAKOV}\vspace{0.7cm}

{\it Bogoliubov Laboratory of Theoretical Physics, Joint
Institute for Nuclear Research, Dubna, Russia \\[0.2cm] and\\[0.2cm]
Institute for Theoretical and Experimental Physics, Moscow,
Russia}
\end{center}

\begin{abstract}
In the present review we show  that renormalizations in a softly
broken SUSY gauge theory are not independent but directly follow
from those of an unbroken or rigid theory. This is a consequence
of a treatment of a softly broken theory as  a rigid one in
external spurion superfield. This enables one to get the singular
part of effective action in a broken theory from a rigid one by a
simple modification of the couplings. Substituting the modified
couplings into renormalization constants, RG equations, solutions
to these equations, approximate solutions, fixed points, etc., one
can get corresponding relations for the soft terms by a simple
Taylor expansion over the Grassmannian variables. Some examples
including the MSSM in low and high $\tan\beta$ regime, SUSY GUTs
and  the N=2 Seiberg-Witten model are considered.
\end{abstract}\vspace{0.3cm}

\section{Introduction}

In a series of papers~\cite{AKK}${}^-$\cite{KV2} we have shown
that renormalizations in a softly broken SUSY theory follow  from
those of an unbroken one   in a  straightforward way. This is in
agreement with the other approaches~\cite{JJ,GR,St} and is
inspired by the original observation of Ref.\cite{Y}. In what
follows we give a review of our approach. It does not explicitly
use the power of holomorphicity advocated by some
authors~\cite{GR,St}, but ends up with the simple and
straightforward algorithm which is easy to apply.

The main idea is that a softly broken supersymmetric gauge theory
can be considered as a rigid SUSY theory imbedded into  external
space-time independent  superfield $\eta$,  so that all couplings
and masses become external superfields $S(\eta,\bar\eta)$. Then,
the following crucial statement is valid~\cite{AKK}

\noindent{\bf The statement:}  {\it In external spurion field
$\eta$ the UV singular part of the effective action depends on the
couplings $S(\eta,\bar\eta)$, but does not depend on their
derivatives:}
\begin{picture}(50,10)(4,36)\put(-140,5){\line(1,1){20}}
\put(-110,5){\line(1,1){20}}\put(-80,5){\line(1,1){20}}
\end{picture}
 \begin{equation}
S^{eff}_{Sing}(g)\Rightarrow S^{eff}_{Sing}(S, D^2S, \bar D^2S,
D^2\bar D^2S),
 \end{equation}
{\it where $D$ and $\bar D$ are the supercovariant derivatives,
and as a result has the same form in unbroken and broken cases.}

With replacement of the couplings by external fields one can
calculate the effective action $S^{eff}_{Sing}(g)$ assuming that
the external field is a constant, i.e. in a rigid theory. This
approach to a softly broken supersymmetric theory allows us to use
remarkable mathematical properties of $N=1$ SUSY field theories
such holomorphicity which leads to the non-renormalization
theorems, cancellation of quadratic divergences, etc.

The renormalization procedure in a softly broken SUSY gauge theory
can be performed in the following way:

 {\it One takes  the renormalization constants of a rigid theory,
calculated in some massless scheme, substitutes instead of the
rigid couplings (gauge and Yukawa) their modified expressions,
which depend on a Grassmannian variable, and expand over this
variable. This gives   renormalization constants for the soft
terms. Differentiating them with respect to a scale one can find
corresponding renormalization group equations.}

 Thus, the soft-term
renormalizations are not independent but can be calculated from
the known renormalizations of a rigid theory with the help of the
differential operators. Explicit form of these operators has been
found in a general case and in some particular models like SUSY
GUTs or the MSSM~\cite{AKK,Kaz}. The same expressions have been
obtained also in a somewhat different approach in
Ref.~\cite{JJ,GR,JJP}.

In fact as it has been shown in \cite{Kaz} this procedure works at all
stages. One can make the above mentioned substitution on the level
of the renormalization constants, RG equations, solutions to these
equations, approximate solutions, fixed points, finiteness
conditions, etc. Expanding then over a Grassmannian variable one
obtains  corresponding expressions for the soft terms. This way
one can get new solutions of  the RG equations and explore their
asymptotics, or approximate solutions, or find their stability
properties, starting from the known expressions for a rigid
theory.

Throughout the paper we assume the existence of some gauge and
SUSY invariant regularization. Though it is   some problem by
itself, in principle it is solvable~\cite{Sl}.  Provided the rigid
theory is well defined, we consider the modifications which appear
due to the presence of soft SUSY breaking terms. To be more
precise, when discussing one, two and three loop calculations of
the renormalization constants we have in mind dimensional
reduction and the minimal subtraction scheme. Though dimensional
reduction is not self-consistent in general~\cite{AV}, it is safe
to use it in low orders and all the actual calculations are
performed in the framework of dimensional
reduction~\cite{Vaughn,Jones,Jack,Ferreira}. Nevertheless, our
main formulae have  general validity provided the invariant
procedure exists.

 Below we give some examples: the general SUSY
gauge theory in higher loops, the MSSM in low $\tan\beta$ regime
where analytical solutions to the one-loop RG equations are known
exactly and in high $\tan\beta$ regime where analytical solutions
are known in iterative or approximate form. We discuss some
particular solutions like the fixed point ones and examine their
properties. The method allows one to get the same type of
solutions for the soft SUSY breaking terms. The other examples are
the  finite N=1 SUSY GUTs and the  N=2 Seiberg-Witten model where
exact (nonperturbative) solution is known. Here one can extend
finiteness and the S-W solution to the soft terms as well.

\section{Soft SUSY Breaking  and  the Spurion Superfields}

Consider an arbitrary $N=1$ SUSY gauge theory with unbroken SUSY
within  the superfield formalism. The Lagrangian of a rigid theory
is given by
\begin{eqnarray}
{\mathcal L}_{rigid} &=& \int d^2\theta~\frac{1}{4g^2}{\rm
Tr}W^{\alpha}W_{\alpha} + \int d^2\bar{\theta}~\frac{1}{4g^2}{\rm
Tr} \bar{W}_{\dot \alpha}\bar{W}^{\dot \alpha} \label{rigidlag} \\
&+& \int d^2\theta d^2\bar{\theta} ~~\bar{\Phi}^i
(e^{V})^j_i\Phi_j + \int
 d^2\theta ~~{\mathcal W} + \int d^2\bar{\theta} ~~\bar{\mathcal W},  \nonumber
\end{eqnarray}
where
 $$W_{\alpha}=-\frac 14\bar D^2e^{-V}D_\alpha e^V,
\ \ \ \bar W_{\dot \alpha}=-\frac 14 D^2e^{-V}\bar D_{\dot \alpha}
e^V,$$ are the gauge field strength tensors and the superpotential
${\mathcal W}$ has the form
\begin{equation}
{\mathcal  W}=\frac{1}{6}y^{ijk}\Phi_i\Phi_j\Phi_k +\frac{1}{2}
M^{ij}\Phi_i\Phi_j.\label{rigidsuppot}
\end{equation}
To fix the gauge, the usual gauge-fixing term can be introduced.
It is useful to choose it in the form
\begin{equation}\label{gf}
{\mathcal L}_{gauge-fixing}= -~\frac{1}{16}\int d^2\theta
d^2\bar{\theta} {\rm Tr}\left(\bar{f}f + f\bar{f}\right)
\end{equation}
where  the gauge fixing condition is taken as
\begin{eqnarray}
f=\bar{D}^2\frac{V}{\sqrt{\xi g^2}}\;,
 ~~~ \bar f = D^2\frac{V}{\sqrt{\xi g^2}}\; . \label{gcond}
\end{eqnarray}
Here $\xi$ is the usual gauge-fixing parameter.
 Then, the corresponding ghost term is \cite{FP}
\begin{equation}\label{ghost}
  {\mathcal L}_{ghost}=i\int d^2\theta~\frac{1}{4}{\rm Tr}~b\,
\delta_{c}f -i\int d^2\bar{\theta}~\frac{1}{4}{\rm Tr}~\bar{b}\,
\delta_{\bar c}\bar f,
\end{equation}
where $c$ and $b$ are the Faddeev--Popov ghost and antighost
chiral superfields, respectively, and $\delta_{c}$ is the gauge
transformation with the replacement of gauge superfield parameters
$\Lambda (\bar \Lambda)$ by chiral (antichiral) ghost fields $c
(\bar c)$.


 For our choice of the gauge-fixing condition,
the gauge transformation of $f$ looks like
\begin{equation}\label{trans}
\delta_{\Lambda} f = \bar D^2\delta_{\Lambda} \frac{V}{\sqrt{\xi
g^2}} = i\bar D^2 \frac{1}{\sqrt{\xi g^2}}{\mathcal
L}_{V/2}[\Lambda+\bar \Lambda + \coth({\mathcal L}_{V/2})(\Lambda
- \bar \Lambda)],
\end{equation}
where ${\mathcal L}_{X}Y\equiv[X,Y]$. Equation (\ref{ghost}) then
takes the form
\begin{eqnarray}
{\mathcal L}_{ghost}&=&- \int d^2\theta~\frac{1}{4}{\rm
Tr}~b\bar{D}^2\frac{1}{\sqrt{\xi g^2}}{\mathcal L}_{V/2}[c+\bar c
+ \coth({\mathcal L}_{V/2})(c - \bar c)] + \hspace{0.3cm} h.c.
\nonumber\\ &=& \int d^2\theta d^2\bar{\theta}~{\rm Tr}~
\left(\frac{b+ \overline{b}}{\sqrt{\xi g^2}}\right) {\mathcal
L}_{V/2}[c+\bar c + \coth({\mathcal L}_{V/2})(c - \bar c)]\label{gh}\\
&&\hspace*{-1.5cm}=\int d^4\theta~{\rm Tr}\left(\frac{b+
\overline{b}}{\sqrt{\xi g^2}}\right) \left(\left
(c-\overline{c}\right) + \frac{1}{2} \Big[V,
\left(c+\overline{c}\right)\Big]+ \frac{1}{12} \bigg[V, \Big[V,
\left(c-\overline{c}\right)\Big]\bigg]+ ... \right). \nonumber
\end{eqnarray}

The resulting Lagrangian together with the gauge-fixing and the
ghost terms are invariant under the BRST transformations.      For
a rigid theory in our normalization of the  fields, they have the
form~\cite{FP}
\begin{eqnarray}
&&\delta V=\epsilon {\mathcal L}_{V/2}[c+\bar c + \coth({\mathcal
L}_{V/2})(c - \bar c)], \nonumber \\
&&\delta\, c^a=-\frac {i}{2}\epsilon f^{abc}c^b c^c\ ,\;\;\;\;\;
\delta\, {\bar c}^a=-\frac {i}{2}\epsilon f^{abc}{\bar c}^b {\bar
c}^c\ ,\nonumber \\ && \delta\, b^a = \frac{1}{8}\epsilon \bar D^2
\bar f^a\ ,  \;\;\;\; \delta\, {\bar b}^a = \frac{1}{8}\epsilon
D^2 f^a. \label{BRSTr}
\end{eqnarray}

Breaking of supersymmetry is the problem by itself. We do not
discuss here the origin of SUSY breaking but rather concentrate on
the consequences of it. Usually one considers the so-called "soft"
breaking of SUSY, which means that the breaking terms do not spoil
renormalizability of the theory and, in particular, the
cancellation of quadratic divergences and are represented by the
operators of dimension less than 4~\cite{GG}. Hence, to perform
the SUSY breaking, that satisfies the requirement of "softness",
one can introduce a gaugino mass term as well as cubic and
quadratic interactions of scalar superpartners of the matter
fields
\begin{eqnarray}\hspace*{-0.8cm}
-{\mathcal L}_{soft-br} &=&\left[ \frac{M}{2}\lambda\lambda +\frac
16 A^{ijk} \phi_i\phi_j\phi_k+ \frac 12 B^{ij}\phi_i\phi_j
+h.c.\right] +(m^2)^i_j\phi^{*}_i\phi^j,\label{sofl}
\end{eqnarray}
where $\lambda$ is  the gaugino field, and $\phi_i$ is the lowest
component of the chiral matter superfield.

This is not the most general form of the soft terms. In principle,
one can add the terms like $\bar \psi\psi, \ \phi^*\phi\phi$,
etc.~\cite{nonst}. However, the conventional choice (\ref{sofl})
is sufficient for the goal of SUSY breaking and in what follows we
stick to it.

Remarkably, one can rewrite  the Lagrangian (\ref{sofl}) in terms
of N=1 superfields introducing  the external spurion
superfields~\cite{GG} $\eta=\theta^2$ and $\bar
\eta=\bar{\theta}^2$, where $\theta$ and $\bar \theta$ are the
Grassmannian parameters, as~\cite{Y}
 \begin{eqnarray} {\mathcal
L}_{soft} &=& \int d^2\theta~\frac{1}{4g^2}(1-2M\theta^2) {\rm
Tr}W^{\alpha}W_{\alpha} + \int
 d^2\bar{\theta}~\frac{1}{4g^2}(1-2\bar{M}\bar{\theta}^2) {\rm
Tr}\bar{W}^{\dot \alpha}\bar{W}_{\dot \alpha}  \nonumber \\
&&+\int d^2\theta d^2\bar{\theta} ~~\bar{\Phi}^i(\delta^k_i
-(m^2)^k_i\eta \bar{\eta})(e^V)^j_k\Phi_j   \label{sofl2} \\ &&+
\int  d^2\theta \left[\frac 16 (y^{ijk}-A^{ijk}
\eta)\Phi_i\Phi_j\Phi_k+ \frac 12 (M^{ij}-B^{ij}\eta )
\Phi_i\Phi_j \right] +h.c. \nonumber
\end{eqnarray}
Thus, one can interpret the soft terms as the modification of the
couplings of a rigid theory. The couplings become external
superfields depending on Grassmannian parameters $\theta$ and
${\bar \theta}$. To get the explicit expression for the modified
couplings, consider eqs.(\ref{sofl2}). The first two terms
give~\cite{AKK}
\begin{equation}
\frac{1}{g^2}\rightarrow\frac{1}{{\tilde g}^2}=\frac{1-M\theta
^2-{\bar M}{\bar \theta}^2}{g^2}. \label{gtilold}
\end{equation}
Since the gauge field strength tensors $W_{\alpha}$ (${\bar
W}_{\alpha}$) are chiral (antichiral) superfields, they enter into
the chiral (antichiral) integrands in eq.(\ref{sofl2}),
respectively. Correspondingly, the $M\theta ^2$ term of
eq.(\ref{gtilold}) contributes to the chiral integral, while the
${\bar M} {\bar \theta}^2$ term contributes to the antichiral one.
There is no $\theta ^2 {\bar \theta}^2$ term in
eq.(\ref{gtilold}), since it is neither chiral, no antichiral and
gives no contribution to eq.(\ref{sofl2}).

We depart here from the holomorphicity arguments~\cite{SV}.
Alternatively one should consider holomorphic and antiholomorphic
gauge couplings and a separate non-chiral superfield to take care
of the mixed term~\cite{GR,St}. This is where different approaches
diverge. It does not lead, however, to any practical difference in
applications within the PT.

Modifying the gauge coupling in the gauge part of the Lagrangian,
one has to do the same in the gauge-fixing (\ref{gcond}) and ghost
(\ref{gh}) parts in order to preserve the BRST invariance. Here
one has the integral over the whole superspace rather than the
chiral one. This means that if one adds to eq.(\ref{gtilold}) a
term proportional to $\theta ^2{\bar{\theta}}^2$, it gives a
nonzero contribution. Moreover, even if this term is not added, it
reappears as a result of renormalization.

We suggest the following modification of eq.(\ref{gtilold})
\begin{equation}
\frac{1}{g^2}\rightarrow\frac{1}{{\tilde g}^2}=\frac{1-M\theta
^2-{\bar M}{\bar \theta}^2-\Delta \theta ^2{\bar{\theta}}^2}{g^2},
\end{equation}
which  gives  the final expression for the soft  gauge
coupling~\cite{KV}
\begin{eqnarray} \tilde{g}^2 = g^2\left(1 + M\theta^2 +
\bar M\bar{\theta}^2 + 2 M {\bar M}\theta^2 \bar{\theta}^2 +
\Delta \theta^2 \bar{\theta}^2\right).  \label{Tildeg}
\end{eqnarray}
 It will be clear below that it is self-consistent to put
$\Delta=0$ in the lowest order of perturbation theory, but it
appears in higher orders due to renormalizations.

One has to take into account, however,  that, since the
gauge-fixing parameter $\xi$ may be considered as an additional
coupling, it also becomes an external superfield and has to be
modified. The soft expression can be written as
\begin{equation}\label{xi}
  \tilde \xi = \xi \left(1+ x\theta^2 + \bar x \bar \theta^2 + (x\bar x
  + z)\theta^2\bar \theta^2\right),
\end{equation}
where parameters $x$ and $z$ can be obtained by solving the
corresponding RG equation (see Appendix B).

Having this in mind, we perform the following modification of the
gauge fixing condition (\ref{gcond}) first used in~\cite{Kond}
\begin{equation}\label{gcond2}
  f \to \bar D^2\frac{V}{\sqrt{\tilde \xi {\tilde g}^2}}, \ \ \ \bar f
  \to D^2\frac{V}{\sqrt{\tilde \xi {\tilde g}^2}}.
\end{equation}
Then, the gauge-fixing term (\ref{gf}) becomes
\begin{equation}\label{gfm}
{\mathcal L}_{gauge-fixing}= -~\frac{1}{8}\int d^2\theta
d^2\bar{\theta} ~{\rm Tr}\left( \bar D^2\frac{V}{\sqrt{\tilde \xi
{\tilde g}^2}} D^2\frac{V}{\sqrt{\tilde \xi {\tilde g}^2}}\right).
\end{equation}
This leads to the corresponding  modification of the associated
ghost term (\ref{ghost})
\begin{eqnarray}\hspace*{-0.5cm}
{\mathcal L}_{ghost} &=&\int d^2\theta d^2\bar{\theta}~{\rm
Tr}~\frac{1}{\sqrt{\tilde \xi \tilde{g}^2}}
\left(b+\overline{b}\right){\mathcal L}_{V/2}[c+\bar c +
\coth({\mathcal L}_{V/2})(c - \bar c)] . \label{ghs}
\end{eqnarray}

To understand the meaning of the $\Delta$ term, consider the
quadratic part of the ghost Lagrangian (\ref{ghs})
\begin{eqnarray}\hspace*{-0.2cm}
{\mathcal L}_{ghost}^{(2)} &= &\int d^4\theta {\rm
Tr}\frac{1}{\sqrt{\xi{g}^2}} \left(1 -\frac 12 M\xi \theta^2 -
\frac 12 {\bar M}\xi \bar{\theta}^2  - \frac 12  \Delta \xi
\theta^2 \bar{\theta}^2\right) (b+\overline{b}) \left
(c-\overline{c}\right)\nonumber \\ &=&\int d^2\theta
d^2\bar{\theta}~{\rm Tr}~\frac{1}{\sqrt{\xi{g}^2}} \left(1 - \frac
12 \Delta \xi \theta^2 \bar{\theta}^2\right) (b+\overline{b})
\left (c-\overline{c}\right)\label{ghs2} \\ &-&\frac 12\int
d^2\theta~{\rm Tr}~\frac{1}{\sqrt{\xi{g}^2}} M\xi  bc + \frac
12\int d^2\bar \theta~{\rm Tr}~\frac{1}{\sqrt{\xi{g}^2}} \bar M
\xi\bar b\bar c ,\nonumber
\end{eqnarray}
where we have used the explicit form of $\tilde \xi$ given in
Appendix B.

If one compares this expression with the usual Lagrangian for the
matter fields (\ref{sofl2}), one finds an obvious identification
of the second line with the soft scalar mass term and the third
line with the mass term in a superpotential. Thus, $\Delta$ plays
the role of a soft mass providing the splitting in the ghost
supermultiplet.

The other place where the $\Delta$-term appears is the
gauge-fixing term (\ref{gfm}). Here it manifests itself as a soft
mass of the auxiliary gauge field, one of the scalar components of
the gauge superfield $V$.

To see this, consider the gauge-fixing term (\ref{gfm}) in more
detail. Expanding the vector superfield $V(x,\theta,\bar \theta)$
in components
\begin{eqnarray*}
V(x, \theta, \bar \theta) & = & {\mathbb C}(x) + i\theta \chi (x)
-i\bar \theta \bar \chi (x) +  \frac{i}{2} \theta \theta N(x) -
\frac{i}{2} \bar
 \theta \bar \theta \bar N(x)
 -  \theta \sigma^{\mu} \bar \theta v_{\mu}(x)\\ &  &\hspace*{-2cm}
  + i \theta \theta
\bar \theta [\bar \lambda (x) + \frac{i}{2}\bar \sigma^{\mu}
\partial _{\mu} \chi (x)]
  -i\bar \theta \bar \theta \theta [\lambda + \frac{i}{2}
\sigma^{\mu} \partial_{\mu} \bar \chi (x)] + \frac{\theta \theta
\bar \theta \bar \theta}{2}[D(x) - \frac{1}{2}\Box {\mathbb
C}(x)].\nonumber
  \end{eqnarray*}
and substituting it into eq.(\ref{gfm}) one finds
\begin{eqnarray}
  {\mathcal L}_{gauge-fixing}&=&\frac{1}{2\xi g^2}\left[-(D-\Box {\mathbb C}
  -\Delta \xi {\mathbb C} +\frac i2
M \xi \bar N -\frac i2 \bar M \xi N)^2 - (\partial ^\mu v_\mu)^2 \right. \nonumber \\
&& \hspace*{-2cm}+\left. (\bar N -i \bar M \xi {\mathbb C})\Box (
N +i M \xi {\mathbb C})-  i(\lambda + \frac 12 \bar M\xi \chi
)\sigma^\mu\partial_\mu (\bar
\lambda + \frac 12 M\xi \bar \chi ) \right. \nonumber \\
&&\hspace*{-2cm}-\left.
 (\lambda + \frac 12 \bar M\xi \chi)\Box \chi - (\bar
\lambda + \frac 12 M\xi \bar \chi )\Box \bar \chi -i\Box
\chi\sigma^\mu \partial_\mu \bar \chi \right].
 \label{com}
\end{eqnarray}
One can see from eq.(\ref{com}) that the parameter $M$, besides
being the gaugino soft mass, plays the role of a mass of the
auxiliary  field $\chi$, while $\Delta$ is the soft mass of the
auxiliary  fields $N$ and ${\mathbb C}$. All these fields are
unphysical degrees of freedom of the gauge superfield. They are
absent in the Wess-Zumino gauge, however, when the gauge-fixing
condition is chosen in supersymmetric form (\ref{gf}), this gauge
is no longer possible, and the auxiliary fields $\chi$, $N$, and
${\mathbb C}$ survive. Thus, the extra $\Delta$ term is associated
with unphysical, ghost, degrees of freedom, just like in the
component approach, one has the mass of unphysical
$\epsilon$-scalars~\cite{eps}. When we go down with energy, all
massive fields decouple, and we get the usual nonsupersymmetric
Yang-Mills theory.

The $\Delta$-term is renormalized and obeys its own RG equation
which can be obtained from the corresponding expression for the
gauge coupling via Grassmannian expansion. In due course of
renormalization, this term is mixing with the soft masses of
scalar superpartners and gives an additional term in RG equations
for the latter.

In component formalism one has a similar term.
 In~\cite{JJP,JJ9709}, the dimensional
reduction (DRED) regularization is used. In this case, one is
bounded to introduce the so-called $\epsilon$-scalars to
compensate the lack of bosonic degrees of freedom in 4-2$\epsilon$
dimensions. These $\epsilon$-scalars in due course of
renormalization acquire a soft mass that enters into the RG
equations for soft masses of physical scalar particles. This
problem has been discussed in~\cite{JJMVY}. If one gets rid of the
$\epsilon$-scalar mass by changing the renormalization scheme,
${\mathrm {DRED}}\rightarrow{\mathrm {DRED'}}$, there appears an
additional term in RG equations for the soft scalar
masses~\cite{MV,JJ9405} called X~\cite{JJP} which usually
coincides with our $\Delta$.

Besides the modification of the gauge coupling the soft terms
(\ref{sofl}) imply the modification of the Yukawa ones and the
mass term
$$y^{ijk}\to\tilde{y}^{ijk}=y^{ijk}-A^{ijk}\eta, \ \ \ M^{ij}
\to\tilde{M}^{ij}=M^{ij}-B^{ij}\eta, \ \ \ +\ \ h.c.$$ There is,
however, also the soft mass term $(m^2)^k_i$.  To take care of it
we absorb the multiplier $(\delta^k_i-(m^2)^k_i\eta\bar\eta)$ into
the redefinition of the fields $\Phi$ and $\bar\Phi$. This results
in the additional modification of the Yukawa couplings and the
mass term
\begin{eqnarray}
y^{ijk}&\to&\tilde{y}^{ijk}=y^{ijk}-A^{ijk}\eta+\frac 12
(y^{njk}(m^2)^i_n
+y^{ink}(m^2)^j_n+y^{ijn}(m^2)^k_n)\eta \bar \eta,\nonumber  \\
M^{ij}& \to&\tilde{M}^{ij}=M^{ij}-B^{ij}\eta+\frac 12
(M^{nk}(m^2)^i_n +M^{in}(m^2)^k_n)\eta \bar \eta.
\end{eqnarray}
This completes our set of substitutions.

At the end of this section, we would like to comment on the BRST
invariance in a softly broken SUSY theory. The BRST
transformations (\ref{BRSTr}) due to our choice of normalization
of the gauge and ghost fields do not depend on the gauge coupling.
Hence, in a softly broken theory they remain unchanged. One can
easily check that, despite the substitution $g^2 \to \tilde g^2$
and $\xi \to \tilde \xi$, the softly broken SUSY theory remains
BRST invariant~\cite{Kond}.

\section{Renormalizations in a Softly Broken SUSY Theory}

The modifications of the couplings introduced above are valid not
only for the classical Lagrangian but also for the quantum one. As
follows from the analysis of the Feynman diagrams in
superspace~\cite{supergraph} the modification of the Feynman rules
due to the soft terms does not influence the UV divergent part of
the effective action~\cite{AKK}. This justifies the statement made
above concerning the UV singular part of the effective action. The
following theorem is valid:

 \noindent {\bf The theorem} {\it Let a rigid theory be
renormalized via introduction of the renormalization constants
{$Z_i$}, defined within some minimal subtraction massless scheme.
Then, a softly broken theory is renormalized via introduction of
the renormalization superfields {$\tilde{Z}_i$} which are related
to {$Z_i$} by the coupling constant redefinition}
 \begin{equation}\label{z}
 {\tilde{Z}_i}(g^2,y ,\bar y)=
Z_i({\tilde{g}^2,\tilde{y},\tilde{\bar y}}),\end{equation}
 {\it
where the redefined couplings are}
 \begin{eqnarray}
\tilde{g}^2_i&=&g^2_i(1+M_i \eta+{\bar M}_i \bar{\eta}+(2 M_i
{\bar M}_i +\Delta_i) \eta \bar{\eta}), \label{g1}\\
\tilde{y}^{ijk}&=&y^{ijk}-A^{ijk}\eta +\frac 12 (y^{njk}(m^2)^i_n
+y^{ink}(m^2)^j_n+y^{ijn}(m^2)^k_n)\eta \bar \eta, \label{y1}\\
\tilde{\bar y}_{ijk}&=&\bar y_{ijk} - \bar A_{ijk} \bar{\eta}+
\frac 12 (y_{njk}(m^2)_i^n +y_{ink}(m^2)_j^n+y_{ijn}(m^2)_k^n)\eta
\bar \eta , \nonumber
\end{eqnarray}

Eqs.(\ref{z}-\ref{y1}) lead to a finite renormalized softly broken
SUSY theory. However, in practice it is more convenient to
consider not the renormalization constants $Z_i$ but the RG
equations directly. Differentiating the renormalization constants
$\tilde{Z}_i$ with respect to a scale one gets the RG functions
for the soft terms of a broken theory in terms of unbroken one.
The resulting soft term $\beta$ functions are summarized below.

\begin{center}
{\bf Summary of the Soft Term Renormalizations }\hspace*{-0.5cm}
\begin{tabular}{|l|l|}
\hline \hline  & \\[-0.2cm]
\hspace*{0.5cm} {The Rigid Terms} & \hspace*{1.5cm}{The Soft
Terms}\\[0.2cm] \hline
$\beta_{\alpha_i} =  \alpha_i\gamma_{\alpha_i}$  & $\beta_{M_{A
i}}=D_1\gamma_{\alpha i}$ \\[0.3cm]
$\beta_{M}^{ij}  =\frac{1}{2}(M^{il}\gamma^j_l+M^{lj}\gamma^i_l)
$& $\beta_{B}^{ij} =
\frac{1}{2}(B^{il}\gamma^j_l+B^{lj}\gamma^i_l)
 - (M^{il}D_1\gamma^j_l+M^{lj}D_1\gamma^i_l) $\\[0.2cm]
$\beta_{y}^{ijk} =\frac{1}{2}(y^{ijl}\gamma^k_l+$ perm's) &
$\beta_{A}^{ijk} = \frac{1}{2}(A^{ijl}\gamma^k_l+$ perm's) -
$(y^{ijl}D_1\gamma^k_l+$ perm's)
   \\[0.2cm]
\hspace*{2cm}$\Uparrow$ &  $(\beta_{m^2})^i_j=D_2\gamma^i_j $\\[0.2cm]
{chiral anomalous dim.}
& $\beta_{\Sigma_{\alpha_i}}=D_2\gamma_{\alpha_i}$\\[0.2cm]
 \hline
\multicolumn{2}{|l|} {}  \\[-0.2cm]
\multicolumn{2}{|l|} {$D_1= M_{A_i}\alpha_i\frac{\displaystyle
\partial}{\displaystyle \partial \alpha_i} -A^{ijk}\frac{\displaystyle
\partial}{\displaystyle \partial y^{ijk}}$, \ \ \ \
$\bar{D}_1=M_{A_i}\alpha_i\frac{\displaystyle
\partial}{\displaystyle
\partial \alpha_i} -A_{ijk}\frac{\displaystyle \partial}{\displaystyle
\partial y_{ijk}}$}
\\[0.2cm]
\multicolumn{2}{|l|} {$D_2= \bar{D}_1 D_1 +
\Sigma_{\alpha_i}\alpha_i\frac{\displaystyle \partial
}{\displaystyle
\partial
\alpha_i}+\frac{1}{2}(m^2)^a_n\left(y^{nbc}\frac{\displaystyle
\partial }{\displaystyle \partial y^{abc}}
+y^{bnc}\frac{\displaystyle \partial
 }{\displaystyle \partial y^{bac}}+
y^{bcn}\frac{\displaystyle \partial }{\displaystyle \partial
y^{bca}}\right. $} \\[0.3cm]
\multicolumn{2}{|l|} {\hspace*{0.1cm} $\left. \ \ \ \ \ \ \ \ \ \
\  + \ y_{abc}\frac{\displaystyle \partial }{\displaystyle
\partial y_{nbc}}+ y_{bac}\frac{\displaystyle \partial
}{\displaystyle \partial y_{bnc}}+ y_{bca}\frac{\displaystyle
\partial
}{\displaystyle \partial y_{bcn}}\right)$,}\\[0.3cm]
\multicolumn{2}{|l|}
{$\Sigma_{\alpha_i}=M_{A_i}\bar M_{A_i} + \Delta_i$} \\[0.3cm]
 \hline \hline
\end{tabular}
\end{center}
Later on we consider some examples of the application of these
formulas.

\section{Grassmannian Taylor Expansion}

We demonstrate now how the RG equations for the soft terms appear
via Grassmannian Taylor expansion from those for the rigid
couplings.

In what follows  we would like to simplify the notations and
consider numerical rather than tensorial couplings. When group
structure and  field content of the model are fixed, one has a set
of  gauge  $\{g_i\}$ and   Yukawa $\{y_k\}$ couplings.  It is
useful to consider the following rigid parameters
 $ \alpha_i \equiv g_i^2/16\pi^2, \  Y_k \equiv
y_k^2/16\pi^2$. Then eqs.(\ref{g1}-\ref{y1})  look like
\begin{eqnarray}
\tilde{\alpha}_i&=&\alpha_i(1+M_i \eta+\bar M_i
\bar{\eta}+(M_i\bar M_i+\Sigma_i)
 \eta \bar{\eta}), \label{ga}\\
\tilde{Y}_k&=&Y_k(1+A_k \eta +\bar A_k \bar{\eta}+ (A_k\bar
A_k+\Sigma_k) \eta \bar \eta),
 \nonumber
\end{eqnarray}
where to standardize the notations we have  redefined parameter A:
$ A \to Ay$  in a usual way  and have changed the sign of A to
match it with the gauge soft terms. Here $\Sigma_k$ stands for a
sum of $m^2$ soft terms, one for each leg in the Yukawa vertex and
$\Sigma_i=M_i\bar M_i+\Delta_i$.

Now the RG equation for a rigid theory can be written in a
universal form
\begin{equation}
\dot  a_i= a_i\gamma_i(a), \ \ \ \   a_i = \{\alpha_i, Y_k\},
\label{RG}
\end{equation}
where $\gamma_i(a)$ stands for a sum of  corresponding anomalous
dimensions. In the same notation the soft terms
(\ref{ga}) take the form
\begin{equation}
\tilde{a}_i=a_i(1+m_i \eta+\bar m_i \bar{\eta}+S_i\eta
\bar{\eta}), \label{usoft}
\end{equation}
where  $m_i=\{M_i, A_k\}$ and  $S_i=\{M_i\bar M_i+\Sigma_i ,
A_k\bar A_k+\Sigma_k \}$.

Substituting  eq.(\ref{usoft}) into eq.(\ref{RG}) and expanding
over $\eta $ and $\bar \eta$ one can get the RG equations for the
soft terms
\begin{equation}
\dot{\tilde{a}}_i= \tilde{a}_i\gamma_i(\tilde{a}), \label{RGs}
\end{equation}
Consider first the F-terms.  Expanding over $\eta$  one has
\begin{equation}
 \dot m_i =  \left.\gamma_i(\tilde{a})\right|_F = \sum_j a_j\frac{\partial
  \gamma_i}{\partial a_j}m_j\equiv D_1\gamma_i.
\end{equation}
This is just the  RG equation for the soft terms $M_i$ and
$A_k$~\cite{JJ,AKK}. Proceeding the same way for the D-terms  and
substituting $S_i= m_i\bar m_i+\Sigma_i$ one has the RG equation
for the mass terms
\begin{equation}
 \dot{\Sigma}_i =  \gamma_i(\tilde{a})\vert_D =
 \sum_ja_j\frac{\partial \gamma_i}{\partial a_j}(m_jm_j+\Sigma_j)  +
  \sum_{j,k} a_ja_k\frac{\partial^2 \gamma_i}{\partial a_j  \partial a_k}m_jm_k
  \equiv D_2\gamma_i.
\end{equation}

One can also obtain  the RG equation for the individual soft
masses out of  field renormalization.  Consider for this purpose
the chiral Green function in a rigid theory. It obeys  the
following RG relation
\begin{equation}
 <\Phi_i \bar{\Phi}_i> \ \ = \ \ <\Phi_i \bar{\Phi}_i>_0 e^{\displaystyle
  \int_0^t \gamma_i(a(t'))dt'}.
\end{equation}
Making the substitution $$<\Phi_i \bar{\Phi}_i> \ \  \to \ \
<\Phi_i \bar{\Phi}_i>(1+m^2_i\eta\bar \eta),
 \ \ \  a\ \to \ \tilde{a},$$
and expanding over $\eta\bar \eta$ ( since it stands under the
full Grassmann
 integral only  D-term contributes) one has
\begin{equation}
  m^2_i=m^2_{i0} + \int_0^t dt' \left. \gamma_i(\tilde{a}(t'))\right|_D .
\end{equation}
 Differentiating this relation with respect to $t$ leads to
 the RG equation for the soft mass
\begin{equation}
  \dot{m^2}_i= D_2 \gamma_i(a).
\end{equation}

\section{Illustration}

Consider, as an illustration of the above formulas, the simplest
case of a pure gauge theory~\cite{Kaz}. In a rigid theory the
coupling is renormalized as
 \begin{equation}
\alpha^{Bare}=Z_{\alpha }\alpha , \ \ \ \ \ \alpha \equiv
g^2/16\pi^2.\end{equation}
 Making the substitution $\alpha \to \tilde\alpha$ one gets \
 ${\tilde{\alpha}^{Bare}=\tilde{Z}_{\alpha }\tilde{\alpha}}
 $ \ or (up to linear terms in $\eta$)
 \begin{equation}
\alpha^{Bare}(1+M_{A}^{Bare}{\eta})=\alpha(1+M_{A}{\eta})
Z_{\alpha }(\alpha(1+M_A\eta)).
 \end{equation}
After expansion over  {$\eta$} this leads to equations
\begin{eqnarray*}
\alpha^{Bare}&=&\alpha Z_{\alpha }(\alpha),  \\
M_{A}^{Bare}\alpha^{Bare}&=&M_{A}\alpha Z_{\alpha }(\alpha)
+\alpha D_1 Z_{\alpha },
\end{eqnarray*}
where $ D_1=M_{A}\alpha\frac{\displaystyle \partial}{\displaystyle
\partial \alpha}$ is the differential operator extracting linear
terms over $\eta$. As a result, we get the bare mass
\begin{equation}
M_{A}^{Bare}=M_{A}+ D_1 \ln Z_{\alpha }. \label{mm}
\end{equation}
Differentiating eq.(\ref{mm}) with respect to a scale, one gets
 \begin{equation}\label{bb}
 \beta_{\alpha}=\alpha \gamma_{\alpha}, \ \ \
\beta_{M_{A}}=D_1\gamma_{\alpha},\end{equation}
 where $\gamma_\alpha$ is the
gauge field anomalous dimension $\gamma_\alpha=-d\log
Z_\alpha/d\log\mu^2 $.

In fact, one does not need eq.(\ref{bb})  to get the RG equation
form the gaugino mass and  can get the same formulas (\ref{bb})
starting directly from the RG equation for $\alpha$ as shown
above. One can go even further and consider a solution to the RG
equation in a rigid theory.

Indeed, let us  take a solution to the RG equation for the
coupling written in quadratures
\begin{equation}\label{al}
  \int^\alpha_{\alpha_0} \frac{d\alpha'}{\beta(\alpha')}  =
  \ln\frac{Q^2}{\mu^2}.
\end{equation}
Performing the replacement of the coupling one gets
\begin{equation} \int^{\alpha(1+M_A\eta+...)}_{\alpha_0(1+M_{A0}\eta+...)}
\frac{d\alpha'}{\beta(\alpha')} = \ln\frac{Q^2}{\mu^2} \label{sin}
\end{equation}
 which
after expansion over $\eta$ leads to the solution for the soft
mass term
 \begin{equation} \frac{\alpha M_A}{\beta(\alpha)}=\frac{\alpha
M_{A0}}{\beta(\alpha_0)} \ \ \Rightarrow \ \ M_A=c_1\
\frac{\displaystyle \beta(\alpha)}{\displaystyle
\alpha}=c_1\gamma(\alpha),\label{mg}
\end{equation} where $\alpha$ is taken
from eq.(\ref{al}). Thus, the solution for the gaugino mass term
directly follows from the one for the rigid coupling.
Eq.(\ref{mg}) is the first example of RG invariants first found
in~\cite{HS} on different grounds. Following our approach one can
construct the other ones~\cite{Kob}. For example, one can continue
the expansion up to D-terms in eq.(\ref{sin}), which gives a
solution for the $\Delta$ term
\begin{equation}
  \Delta=c_2 \gamma(\alpha) - c_1\alpha \gamma'(\alpha)\gamma(\alpha).
\end{equation}

\section{Examples}

\subsection{General gauge theory }

In the one-loop order the rigid $\beta$ functions are (for
simplicity, we consider the case of a single gauge coupling)
\begin{eqnarray}
\beta_{\alpha} &=&  \alpha\gamma_{\alpha}, \ \
\gamma_\alpha^{(1)}=\alpha Q, \ \ \ Q=T(R)-3C(G), \\
 \beta_{y}^{ijk}&
=&\frac{1}{2}(y^{ijl}\gamma^k_l+ perm's),\ \  \gamma^{i\ (1)}_j=
\frac{1}{2}y^{ikl}y_{jkl}-2\alpha C(R)^i_j,\nonumber
\end{eqnarray}
where $T(R),C(G)$ and $C(R)$ are the Casimir operators of the
gauge group  defined by
$$T(R)\delta_{AB}=Tr(R_AR_B), \ \ C(G)\delta_{AB}=f_{ACD}f_{BCD}, \ \
C(R)^i_j=(R_AR_A)^i_j .$$

Applying our algorithm, this leads to the following soft $\beta$
functions:
\begin{eqnarray}
\beta_{M_A}^{(1)} &=& \alpha M_AQ, \\
\beta_B^{ij\ (1)} &=& \frac{1}{2}B^{il}(\frac{1}{2}y^{jkm}y_{lkm}
-2\alpha C(R)^j_l) \\ &+& M^{il}(\frac{1}{2}A^{jkm}y_{lkm}+2\alpha
M_AC(R)^j_l)
+(i\leftrightarrow j), \nonumber \\
\beta_{A}^{ijk\
(1)}&=&\frac{1}{2}A^{ijl}(\frac{1}{2}y^{kmn}y_{lmn} -2\alpha
C(R)^k_l) \\ &+& y^{ijl}(\frac{1}{2}A^{kmn}y_{lmn}+2\alpha
M_AC(R)^k_l)
 +(i\leftrightarrow j,k), \nonumber\\
\left[\beta_{m^2}\right]^{i\ (1)}_j &=&
\frac{1}{2}A^{ikl}A_{jkl}-4\alpha M_A^2C(R)^i_j \\ &+&
\frac{1}{4}y^{nkl}(m^2)^i_ny_{jkl}+\frac{1}{4}y^{ikl}(m^2)^n_jy_{nkl}
+\frac{4}{4}y^{isl}(m^2)^k_sy_{jkl}.\nonumber
\end{eqnarray}
We used here the fact that in the given order the solution for
$\Sigma_\alpha$ is $\Sigma_\alpha=M_A\bar M_A$.

In two loops the  rigid  anomalous dimensions are
\begin{eqnarray*}
\gamma_\alpha^{(2)}&=&2\alpha^2C(G)Q-\frac{2\alpha }{r}C(R)^i_j
(\frac{1}{2}y^{jkl}y_{ikl}-2\alpha C(R)^j_i), \ \ r=dim
G=\delta_{AA},
\\ \gamma^{i\ (2)}_j&=&-(y^{imp}y_{jmn}+2\alpha C(R)^p_j\delta^i_n)
(\frac{1}{2}y^{nkl}y_{pkl}-2\alpha C(R)^n_p)+2\alpha^2QC(R)^i_j.
\end{eqnarray*}

In this case, again the solution for the ghost mass
$\Delta_\alpha$ can be found analytically~\cite{KV} and coincides
with  the mass of $\epsilon$-scalars~\cite{JJ9803}
\begin{eqnarray}
 {\Sigma_{\alpha}}^{(2)}&=& \Delta_{\alpha}^{(2)}=
-2\alpha [ \frac{1}{r} {(m^2)}_j ^i {C(R)}_i ^j - M^2_A C(G)].
\label{sigma2}
\end{eqnarray}

Then, the soft renormalizations are as follows:
\begin{eqnarray}
\beta_{M_A}^{(2)} &=& 4\alpha^2M_AC(G)Q-\frac{2\alpha
M_A}{r}C(R)^i_j
(\frac{1}{2}y^{jkl}y_{ikl}-2\alpha C(R)^j_i) \nonumber \\
&&+\frac{2\alpha}{r}C(R)^i_j(\frac{1}{2}A^{jkl}y_{ikl}+2\alpha
M_AC(R)^j_i), \\
\beta_B^{ij\ (2)} &=& -\frac{1}{2}B^{il}(y^{jkp}y_{lkn}+2\alpha
C(R)^p_l\delta^j_n)(\frac{1}{2}y^{nst}y_{pst}-2\alpha C(R)^n_p)\nonumber \\
&& -M^{il}(A^{jkp}y_{lkn}-2\alpha M_AC(R)^p_l\delta^j_n)
(\frac{1}{2}y^{nst}y_{pst}-2\alpha C(R)^n_p) \nonumber \\
&& -M^{il}(y^{jkp}y_{lkn}+2\alpha C(R)^p_l\delta^j_n)
(\frac{1}{2}A^{nst}y_{pst}+2\alpha M_AC(R)^n_p) \nonumber \\
&&+B^{il}\alpha^2QC(R)^j_l-4M^{il}\alpha^2QC(R)^j_lM_A +
(i\leftrightarrow j), \\
\beta_{A}^{ijk\ (2)} &=&
-\frac{1}{2}A^{ijl}(y^{kmp}y_{lmn}+2\alpha
C(R)^p_l\delta^k_n)(\frac{1}{2}y^{nst}y_{pst}-2\alpha C(R)^n_p)\nonumber \\
&& +A^{ijl}\alpha^2QC(R)^k_l-4y^{ijl}\alpha^2QC(R)^j_lM_A\nonumber \\
&&-y^{ijl}(A^{kmp}y_{lmn}-2\alpha M_AC(R)^p_l\delta^k_n)
(\frac{1}{2}y^{nst}y_{pst}-2\alpha C(R)^n_p) \nonumber \\
&& -y^{ijl}(y^{kmp}y_{lmn}+2\alpha C(R)^p_l\delta^k_n)
(\frac{1}{2}A^{nst}y_{pst}+2\alpha M_AC(R)^n_p) \nonumber \\
&&+ (i\leftrightarrow j)+ (i\leftrightarrow k),\\
\left[\beta_{m^2}\right]^{i\ (2)}_j
&=&-(A^{ikp}A_{jkn}+\frac{1}{2}(m^2)^i_ly^{lkp}y_{jkn}
+\frac{1}{2}y^{ikp}y_{lkn}(m^2)^l_j \nonumber \\
&&
+\frac{2}{2}y^{ilp}(m^2)^s_ly_{jsn}+\frac{1}{2}y^{iks}(m^2)^p_sy_{jkn}+
\frac{1}{2}y_{ikp}(m^2)^s_ny_{jks} \nonumber \\
&&+4\alpha
M_A^2C(R)^p_j\delta^i_n)(\frac{1}{2}y^{nst}y_{pst}-2\alpha
C(R)^n_p) \nonumber \\
&&-(y^{ikp}y_{jkn}+2\alpha
C(R)^p_j\delta^i_n)(\frac{1}{2}A^{nst}A_{pst}
+\frac{1}{4}(m^2)^k_ly^{lst}y_{pst} \nonumber \\
&&+\frac{1}{4}y^{nst}y_{lst}(m^2)^l_p+\frac{4}{4}y^{nlt}(m^2)^s_ly_{pst}
-2\alpha M_A^2C(R)^n_p )\\
&&-(A^{ikp}y_{jkn}-2\alpha M_AC(R)^p_j\delta^i_n)
(\frac{1}{2}y^{nst}A_{pst}+2\alpha M_AC(R)^n_p) \nonumber \\
&&-(y^{ikp}A_{jkn}-2\alpha M_AC(R)^p_j\delta^i_n)
(\frac{1}{2}A^{nst}y_{pst}+2\alpha M_AC(R)^n_p) \nonumber \\
&& + 12\alpha^2M_A^2QC(R)^i_j +4\alpha^2 C(R)^i_j [ \frac{1}{r}
{(m^2)}^k_l {C(R)}^l_k - M^2 C(G)]\nonumber ,
\end{eqnarray}
where the last term is an extra contribution due to nonzero
$\Delta^{(2)}_i$ in (\ref{sigma2}).

To demonstrate the power of the proposed algorithm, we calculate
the three loop gaugino mass renormalization out of a gauge $\beta
$ function.  One has in three loops~\cite{Jack}
\begin{eqnarray}
\gamma_\alpha^{(3)}&=&\alpha^3C(G)Q[4C(G)-Q]-\frac{6}{r}
\alpha^3QC(R)^i_jC(R)^j_i \nonumber \\
&+&\frac{3}{r}\alpha^2(y^{ikl}y_{jkl}-4\alpha
C(R)^i_j)C(R)^j_sC(R)^s_i 
- \frac{2}{r}\alpha^2C(G)(y^{ikl}y_{jkl}\nonumber \\&-&4\alpha
C(R)^i_j)C(R)^j_i +\frac{3}{2r}\alpha
y^{ikm}y_{jkn}(y^{nst}y_{mst}-4\alpha
C(R)^n_m)C(R)^j_i \nonumber \\
&+&\frac{1}{4r}\alpha (y^{ikl}y_{jkl}-4\alpha C(R)^i_j)
(y^{jst}y_{pst}-4\alpha C(R)^j_p)C(R)^p_i .
\end{eqnarray}
The corresponding gaugino mass renormalization is
\begin{eqnarray} \beta_{M_A}^{(3)} &=&
3\alpha^3C(G)Q[4C(G)-Q]M_A-\frac{18}{r}
\alpha^3QC(R)^i_jC(R)^j_iM_A \nonumber \\
&&\hspace*{-0.6cm}+\ \frac{6}{r}\alpha^2(y^{ikl}y_{jkl}-4\alpha
C(R)^i_j)C(R)^j_sC(R)^s_iM_A
-\frac{3}{r}\alpha^2(A^{ikl}y_{jkl}\nonumber
\\&&\hspace*{-0.6cm}+\ 4\alpha C(R)^i_jM_A) C(R)^j_sC(R)^s_i
-\frac{4}{r}\alpha^2C(G)(y^{ikl}y_{jkl}-4\alpha C(R)^i_j)C(R)^j_i
M_A \nonumber \\
&&\hspace*{-0.6cm}+\
\frac{2}{r}\alpha^2C(G)(A^{ikl}y_{jkl}+4\alpha
C(R)^i_jM_A)C(R)^j_i
+\frac{3}{2r}\alpha y^{ikm}y_{jkn}(y^{nst}y_{mst}\nonumber\\
&&\hspace*{-0.6cm} -\ 4\alpha C(R)^n_m)C(R)^j_i M_A
-\frac{3}{2r}\alpha A^{ikm}y_{jkn}(y^{nst}y_{mst}-4\alpha
C(R)^n_m)C(R)^j_i \nonumber \\
&&\hspace*{-0.6cm}-\ \frac{3}{2r}\alpha
y^{ikm}y_{jkn}(A^{nst}y_{mst}+4\alpha
C(R)^n_mM_A)C(R)^j_i \nonumber \\
&&\hspace*{-0.6cm}+\ \frac{1}{4r}\alpha (y^{ikl}y_{jkl}-4\alpha
C(R)^i_j)
(y^{jst}y_{pst}-4\alpha C(R)^j_p)C(R)^p_i M_A \nonumber \\
&&\hspace*{-0.6cm}-\ \frac{1}{4r}\alpha (A^{ikl}y_{jkl}+4\alpha
C(R)^i_jM_A)
(y^{jst}y_{pst}-4\alpha C(R)^j_p)C(R)^p_i  \nonumber \\
&&\hspace*{-0.6cm}-\ \frac{1}{4r}\alpha (y^{ikl}y_{jkl}-4\alpha
C(R)^i_j) (A^{jst}y_{pst}+4\alpha C(R)^j_pM_A)C(R)^p_i.
\end{eqnarray}

To argue that a solution for $\Delta_i$ exists in all orders of
PT, one can consider the so-called NSVZ-scheme \cite{NSVZ} where
the anomalous dimension $\gamma_\alpha$  is known to all orders of
PT
\begin{equation}\label{NSVZ}
  \gamma_\alpha^{NSVZ}= \alpha \frac{\displaystyle Q-2r^{-1}
   {\rm Tr}[\gamma C(R)]}{\displaystyle 1-2C(G)\alpha}.
\end{equation}
Then the solution for  $\Delta_\alpha$  is
\begin{equation}
  \Delta_\alpha^{NSVZ}=-2 \alpha \frac{\displaystyle
   r^{-1} {\rm Tr}[m^2 C(R)]-M_A^2C(G)}{\displaystyle 1-2C(G)\alpha}.
\end{equation}
and coincides with the for $\epsilon$-scalar
mass~\cite{JJ9803,KKZ}.

\subsection{The MSSM in low $\tan\beta$ regime}

Consider the MSSM in low $\tan\beta$ regime.  One has three gauge
and one Yukawa
 coupling. The one-loop RG equations are~\cite{Ibanez}
\begin{eqnarray}
\dot{\alpha}_i&=&-b_i\alpha^2_i, \ \ \ \  b_i=(\frac{33}{5},1,-3),
\ \ i=1,2,3, \\
\dot{Y}_t&=&Y_t(\frac{16}{3}\alpha_3+3\alpha_2+\frac{13}{15}\alpha_1-6Y_t),
\end{eqnarray}
with the initial conditions: $\alpha_i(0)=\alpha_0, \ Y_t(0)=Y_0$
and $t=\ln(M_X^2/Q^2)$.
 Their solutions are given by~\cite{Ibanez}
\begin{equation}
\alpha_i(t)=\frac{\alpha_0}{1+b_i\alpha_0t}, \ \ \
Y_t(t)=\frac{Y_0E(t)}{1+6Y_0F(t)},
 \label{sol}
\end{equation}
where
\begin{eqnarray*}
E(t)&=&\prod_i(1+b_i\alpha_0t)^{\displaystyle c_i/b_i} , \ \ \
 c_i=(\frac{13}{15},3,\frac{16}{3}),\ \ \
F(t)=\int^t_0 E(t')dt'.
\end{eqnarray*}

To get the solutions for the soft terms it is enough to perform
the substitution $\alpha \to \tilde{\alpha}$ and $Y\to \tilde{Y}$
and expand over $\eta $ and $\bar \eta$. Expanding the gauge
coupling in (\ref{sol}) up to $\eta$ one has (hereafter we assume
$M_{i0}=M_0$)
$$\alpha_iM_i=\frac{\alpha_0M_0}{1+b_i\alpha_0t}-\frac{\alpha_0
b_i\alpha_0M_0t}{(1+b_i\alpha_0t)^2}=\frac{\alpha_0}{1+b_i\alpha_0t}\
. \ \frac{M_0}{1+b_i\alpha_0t},$$ or
\begin{equation}
M_i(t)=\frac{M_0}{1+b_i\alpha_0t}.
\end{equation}
Performing the same expansion for the Yukawa coupling and using
the relations $$\left.
\frac{d\tilde{E}}{d\eta}\right|_\eta=M_0t\frac{dE}{dt}, \ \ \
\left.\frac{d\tilde{F}}{d\eta}\right|_\eta=M_0(tE-F),$$ one finds
a well known expression~\cite{Ibanez}
\begin{equation}
A_t(t)=\frac{A_0}{1+6Y_0F}+M_0\left( \frac{t}{E}\frac{dE}{dt}-
\frac{6Y_0}{1+6Y_0F}(tE-F)  \right). \label{a}
\end{equation}
To get the solution for the term
$\Sigma_t=\tilde{m}^2_t+\tilde{m}^2_Q+m^2_{H_2}$  one has to make
expansion over $\eta$ and $\bar \eta$. This can be done with the
help of the following relations
 $$\left.\frac{d^2\tilde{E}}{d\eta
d\bar \eta}\right|_{\eta,\bar \eta}=
M_0^2\frac{d}{dt}\left(t^2\frac{dE}{dt}\right), \ \ \
\left.\frac{d^2\tilde{F}}{d\eta d\bar \eta}\right|_{\eta,\bar
\eta} =M_0^2t^2\frac{dE}{dt},$$ and leads to~\cite{Kaz}

\begin{eqnarray}
\Sigma_t(t)&=&\frac{\Sigma_0-A_0^2}{1+6Y_0F}
+\frac{(A_0-M_06Y_0(tE-F))^2}{(1+6Y_0F)^2}\label{si}\\
&&+M_0^2\left[\frac{d}{dt}\left(\frac{t^2}{E}\frac{dE}{dt}\right)
-\frac{6Y_0}{1+6Y_0F}t^2\frac{dE}{dt}\right].\nonumber
\end{eqnarray}

With  analytic solutions (\ref{a},\ref{si}) one can analyze
asymptotics and, in particular, find the so-called infrared quasi
fixed points~\cite{Hill} which correspond to $Y_0 \to \infty$
\begin{eqnarray}
Y^{FP}_t&=&\frac{E}{6F},  \label{Yf}\\
A^{FP}_t&=&M_0\left(\frac{t}{E}\frac{dE}{dt}-\frac{tE-F}{F}\right),
\label{Af}\\
\Sigma^{FP}_t&=&M_0^2\left[\left(\frac{tE-F}{F}\right)^2+\frac{d}{dt}
\left(\frac{t^2}{E}\frac{dE}{dt}\right)-\frac{t^2}{F}\frac{dE}{dt}\right].
\label{Sf}
\end{eqnarray}
However, the advantage of the Grassmannian expansion procedure is
that one can perform it for  fixed points as well. Thus the FP
solutions (\ref{Af},\ref{Sf}) can be directly obtained from a
fixed point for the rigid Yukawa coupling (\ref{Yf}) by
Grassmannian expansion. This explains, in particular, why fixed
point solutions for the soft couplings  exist if they exist for
the rigid ones and with the same stability properties~\cite{JJst}.

\subsection{The MSSM in high $\tan\beta$ regime}

Consider the MSSM in high $\tan\beta$ regime.  One has three gauge
and three Yukawa couplings. The one-loop RG equations
are~\cite{Ibanez}
\begin{eqnarray}
\dot{\alpha}_i = -b_i\alpha_i^2, \ \ \ \ \dot{Y}_k =
Y_k(\sum_{i}c_{ki}\alpha_i - \sum_{l}a_{kl}Y_l), \label{y}
\end{eqnarray}
where $i=1,2,3;\ k=t,b,\tau$, $\cdot \equiv d/dt, \ t= \log
M_{GUT}^2/Q^2$ and
\begin{eqnarray*}
b_i&=&\{33/5,1,-3 \},\ \ \ \ a_{tl}=\{6,1,0 \},\ \ a_{bl}=\{1,6,1
\},\ \ a_{\tau l}=\{0,3,4 \},
 \\ c_{ti}&=& \{13/15,3,16/3 \}, \ \
c_{bi}=\{7/15,3,1 6/3 \},\ \  c_{\tau i}=\{9/5,3,0 \}.
\end{eqnarray*}

 Despite a simple form of these equations, there is no explicit
analytic solution similar to (\ref{sol}). One has either the
iterative solution~\cite{AM} or the approximate one~\cite{CK}. In
both the cases the Grassmannian expansion over $\eta$ leads to the
corresponding solutions for the soft terms.

Consider first the iterative solution. It can be written as
\cite{AM}
\begin{equation}
\alpha_i= \frac{\alpha_i^0}{1+b_i\alpha_i^0t}, \ \ \ \  Y_k  =
\frac{Y_k^0u_k}{1+a_{kk}Y_k^0\int_0^t u_k},
 \label{soly}
\end{equation}
where the functions $\{ u_k\}$ obey the integral system of
equations
{\small
\begin{eqnarray}
u_t & = & \frac{E_t}{(1+6Y_b^0\int_0^t u_b)^{1/6}}, \ \ \ \
u_\tau \; = \;
\frac{E_\tau}{(1+6Y_b^0\int_0^t u_b)^{1/2}}
,\nonumber \\
u_b & =&  \frac{E_b}{(1+6Y_t^0\int_0^t u_t)^{1/6}
(1+4Y_\tau^0\int_0^t u_\tau)^{1/4}}
, \label{u} \end{eqnarray}}
and the functions $E_k$ are given by $ E_k=
\prod_{i=1}^3(1+b_i\alpha_i^0t)^{c_{ki}/b_i}.$

 Let us stress that
eqs.(\ref{soly}) give the exact solution to
eqs.(\ref{y}), while the $u_k$'s in eqs.(\ref{u}),
although solved formally in terms of the $E_k$'s and $Y_k^0$'s as
continued integrated fractions, should in practice be solved
iteratively.

To get the solutions for the soft terms it is enough to perform
substitution $\alpha_i \to \tilde{\alpha}_i$ and $Y_k\to
\tilde{Y}_k$ and expand over $\eta $ and $\bar \eta$. One has
\cite{KM}:
\begin{eqnarray}
M_i= \frac{M_i^0}{1+b_i\alpha_i^0t}, \ \ \ \ A_k = -e_k +
\frac{A_k^0/Y_k^0 +a_{kk}\int u_ke_k}{1/Y_k^0 +a_{kk}\int
u_k},\label{A} \nonumber \\ \Sigma_k = \xi_k+A_k^2+2e_kA_k
-\frac{(A_k^0)^2/Y_k^0 -\Sigma_k^0/Y_k^0+a_{kk}\int
u_k\xi_k}{1/Y_k^0+a_{kk}\int u_k},
\end{eqnarray}
where the new functions $e_k$ and $\xi_k$ have been introduced
which obey the iteration equations. For illustration we present
below the corresponding expressions for $e_t$ and $\xi_t$
\begin{eqnarray}
 e_t &=& \frac{1}{E_t}\frac{d\tilde{E}_t}{d\eta}+ \frac{A_b^0\int
u_b-\int u_be_b}{ 1/{Y_b^0}+6\int u_b } ,
\nonumber \\ &&   \nonumber \\ \xi_t &=&
\frac{1}{E_t}\frac{d^2\tilde{E}_t}{d\eta d\bar\eta}
 +2\frac{1}{E_t}\frac{d\tilde{E}_t}{d\eta} \frac{A_b^0\int
u_b-\int u_be_b}{ {1}/{Y_b^0}+6\int u_b }
+ 7 \frac{\left(A_b^0\int u_b-\int u_be_b\right)^2}{\left(
{1}/{Y_b^0} +6\int u_b \right)^2}  \nonumber
\\
 &&-\frac{(\Sigma_b^0+(A_b^0)^2)\int u_b
-2A_b^0\int u_be_b +\int u_b\xi_b}{{1}/{Y_b^0}
+6\int u_b }, \label{ex}
\end{eqnarray}
where the variations of $\tilde{E}_k$ should be taken at $\eta =
\bar \eta=0$ and are given by
\begin{eqnarray*}
\left.\frac{1}{E_k}\frac{d\tilde{E}_k}{d\eta}\right|_{\eta,\bar
\eta=0}&&\hspace*{-0.3cm}= t\sum_{i=1}^3c_{ki}\alpha_iM_i^0, \label{var1}\\
\left.\frac{1}{E_k}\frac{d^2\tilde{E}_k}{d\eta
d\bar\eta}\right|_{\eta,\bar \eta=0}&&\hspace*{-0.7cm}=
t^2\left(\sum_{i=1}^3c_{ki}\alpha_iM_i^0 \right)^2 +2t
\sum_{i=1}^3c_{ki}\alpha_i(M_i^0)^2 -t^2
\sum_{i=1}^3c_{ki}b_i\alpha_i^2(M_i^0)^2. \label{var2}
\end{eqnarray*}
When solving eqs.(\ref{u}) and (\ref{ex}) in the $n$-th iteration
one has to substitute in the r.h.s. the $(n-1)$-th iterative
solution for all the corresponding functions.

The same procedure works for the approximate
 solutions. Once one gets an approximate solution for
 the Yukawa couplings, one immediately has those  for the soft terms
as well \cite{CK}.

We  consider as an illustration the approximate solution. It can
be taken in the form~\cite{CK}
\begin{eqnarray}\hspace*{-1cm}
Y_t^{app}(t)&=&\frac{Y_{t0}E_t(t)}{[1+\frac
72(Y_{t0}F_t(t)+Y_{b0}F_b(t))]^{2/7} [1+7Y_{t0}F_t(t)]^{5/7}}\nonumber\\
Y_b^{app}(t)&=&\frac{Y_{b0}E_b(t)}{[1+\frac
72(Y_{t0}F_t(t)+Y_{b0}F_b(t))]^{2/7}
[1+7Y_{t0}F_t(t)]^{2/7}},\label{app} \\&&\hspace*{1cm} \times \
\frac{1}{[1+\frac
73(3Y_{b0}F_b(t)+Y_{\tau 0}F_\tau)]^{3/7}},\nonumber\\
Y_\tau^{app}(t)&=&\frac{Y_{\tau0}E_\tau(t)}{[1+\frac{21}{4}Y_{\tau
0}F_\tau]^{4/7} [1+\frac 73(3Y_{b0}F_b(t)+Y_{\tau
0}F_\tau)]^{3/7}}.\nonumber
\end{eqnarray}

To demonstrate the accuracy of the approximate solution
(\ref{app}) and the efficiency of the Grassmannian expansion, we
present in Fig.1 the comparison of numerical and approximate
solutions for the Yukawa couplings of a rigid theory as well as
the soft terms.
\begin{figure}[ht]
\hspace*{-0.2cm}\epsfxsize=7.3cm
\epsffile{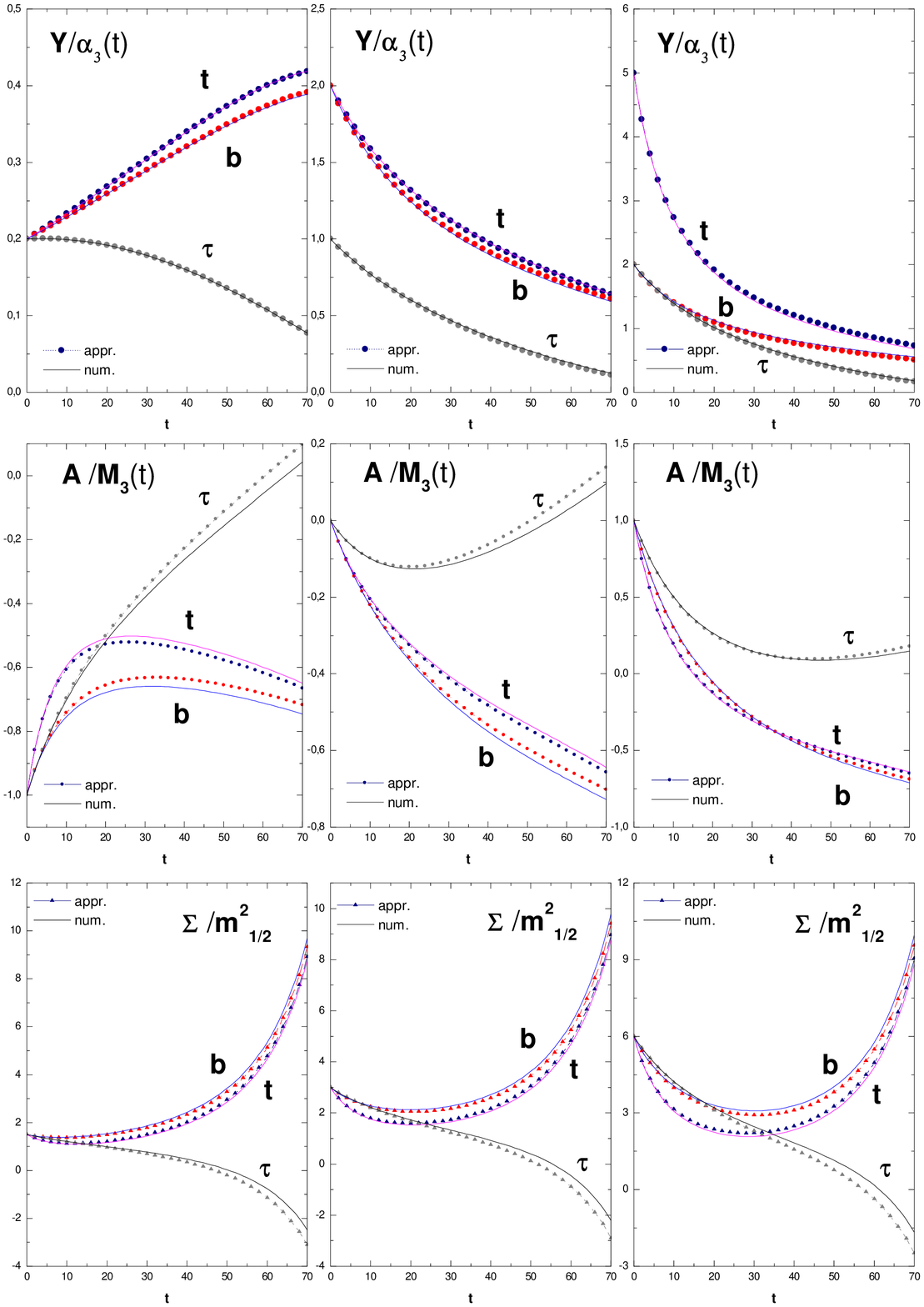}\hspace{0.1cm}
 \epsfxsize=7.3cm \epsffile{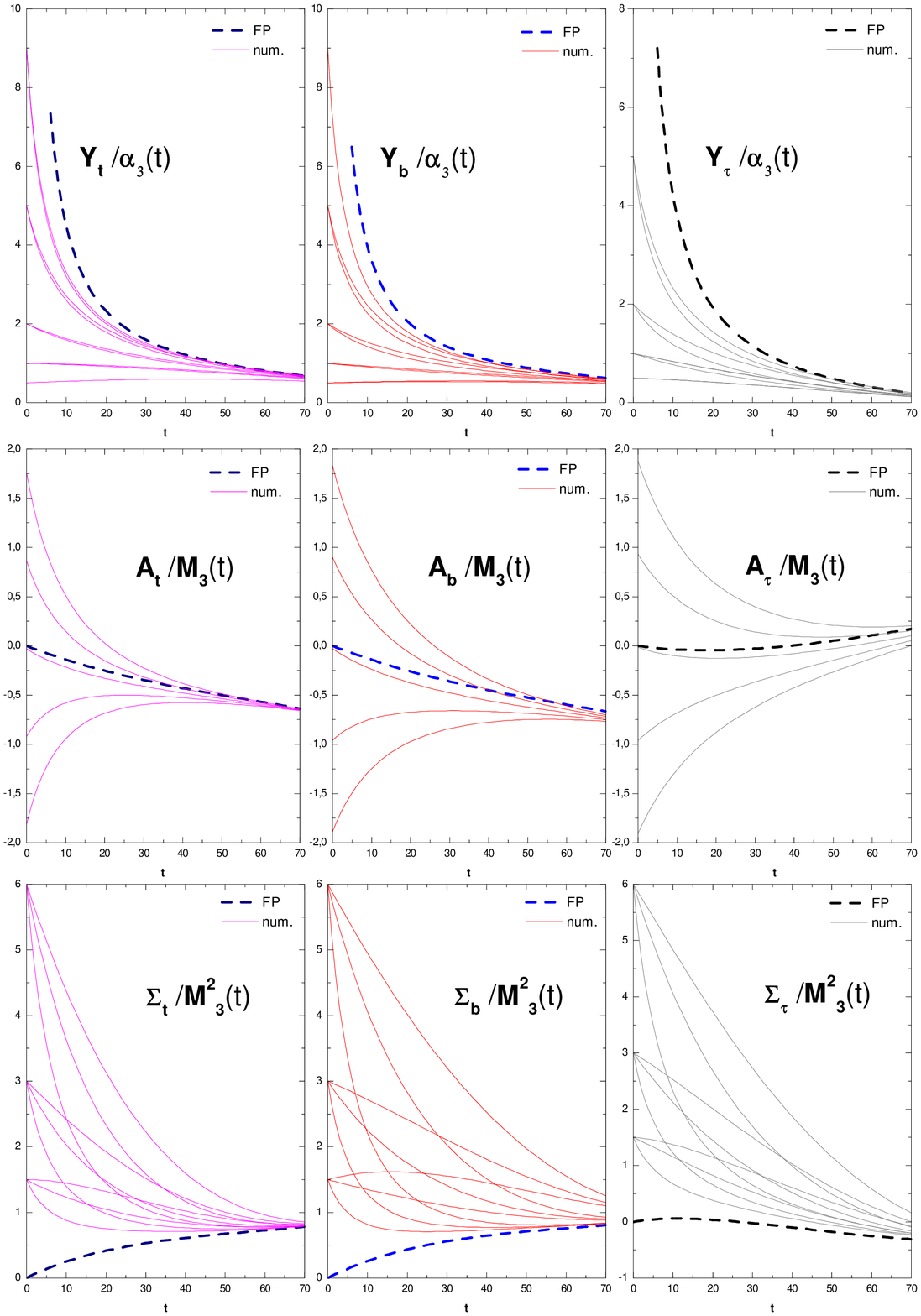}
 \caption{Comparison of numerical and approximate solutions.
 Dotted lines correspond to the analytical approximate solutions,
 solid lines to the numerical solution. Shown also are the
 infra-red quasi fixed points obtained via Grassmannian expansion
 and the numerical curves approaching them in the IR limit.
  }
\end{figure}

One can notice perfect agreement of numerical and analytical
curves. Shown also are the fixed point behaviour, again for the
Yukawa couplings  and for the soft terms obtained via the
expansion procedure for the approximate solutions (\ref{app}). The
numerical curves approach the analytically calculated FP's in the
infrared region.

\subsection{Totally all loop finite N=1 SUSY gauge theories}
Another example of application of the same procedure is the
so-called finite field theories in the framework of SUSY GUTs.
These are the theories where all the UV divergences cancel and
hence all the $\beta$ functions vanish. This can be achieved in a
rigid theory if the following two conditions are
satisfied~\cite{Finite,EKT}:

 $\bullet$ The group representations
are chosen in a way to obey the sum rule
\begin{equation}\sum T(R)=3C_2(G)\end{equation}

 $\bullet$ The Yukawa couplings
are the functions of the gauge one
\begin{equation}\label{fin}
Y_i=Y_i(\alpha),\ Y_i(\alpha)=c^i_1\alpha +c^i_2\alpha^2+...
\end{equation}
where the coefficients $c^i_n$  are calculated algebraically in
the n-th order of PT.

To achieve the complete finiteness of the model including the soft
terms, one has to modify the finiteness condition (\ref{fin}) as
\begin{equation} \tilde Y_i = Y_i(\tilde \alpha)\end{equation}
and perform the expansion over $\eta,\bar\eta$. This
gives~\cite{K} \begin{equation} \left\{\begin{array}{l}
A_i=-M_A\frac{\displaystyle d\ln
Y_i}{\displaystyle d\ln\alpha},\\ \\
\Sigma_i=M^2_A\frac{\displaystyle d}{\displaystyle
d\alpha}\alpha\frac{\displaystyle d\ln Y_i}{\displaystyle
d\ln\alpha},
\end{array}\right.\end{equation}
where $Y(\alpha)$ is assumed to be known from a rigid theory.
 These expressions lead to a totally finite softly broken SUSY
field theory!

Alternatively one can formulate the same conditions in terms of
the bare couplings. They are finite in this case. In dimensional
regularization one has instead of eq.(\ref{fin})
 \begin{equation}\label{d} Y_i^{Bare}=\alpha_{Bare}\cdot
f_i(\varepsilon), \ \ \
f_i(\varepsilon)=c_i^{(1)}+c_i^{(2)}\varepsilon+c_i^{(3)}\varepsilon^2+...\end{equation}
where the coefficients $c_i^{(n)}$ are in one-to-one
correspondence to those in eq.(\ref{fin}). Replacing the couplings
in eq.(\ref{d}) in a usual way one finds that the function
$f(\varepsilon)$ cancels and  one has simple relations for the
soft terms valid in all orders of PT~\cite{K}
 \begin{equation} \tilde Y_i^{Bare}=\tilde\alpha_{Bare}\cdot f_i(\varepsilon) \
\Rightarrow \ \left\{
\begin{array}{l} A_i^{Bare}=-M^{Bare}_A,\\
\Sigma_i^{Bare}=(M^{Bare}_A)^2.\end{array}\right.\end{equation}
These relations for the bare quantities provide the vanishing of
the $\beta$ functions for the soft terms.

\subsection{N=2 SUSY Seiberg-Witten Theory}

Consider now the N=2 supersymmetric gauge theory. The Lagrangian
written in terms of N=2 superfields is~\cite{AG}:
\begin{equation}
{\mathcal L}=\frac{1}{4\pi}{\mathcal I}m Tr \int d^2\theta_1
d^2\theta_2
 \frac 12 \tau \Psi^2,
\end{equation}
where N=2 chiral superfield $\Psi(y,\theta_1,\theta_2)$ is defined
by constraints $\bar{D}_{\dot{\alpha}}\Psi=0$ and
$\bar{\tilde{D}}_{\dot{\alpha}}\Psi=0$
 and
\begin{equation}
\tau = i\frac{4\pi}{g^2}+\frac{\theta}{2\pi}^{\hspace{-0.1cm}
topological}\hspace{-1.1cm}.
\end{equation}

The expansion of $\Psi$  in terms of $\theta_2$  can be written as
$$\Psi(y,\theta_1,\theta_2)=\Psi^{(1)}(y,\theta_1)+\sqrt{2}\theta_2^\alpha
 \Psi^{(2)}_\alpha(y,\theta_1)+\theta_2^\alpha \theta_{2\alpha} \Psi^{(3)}
 (y,\theta_1),$$
where $y^\mu=x^\mu+i\theta_1\sigma^\mu\bar \theta_1 +i\theta_2
\sigma^\mu\bar\theta_2$  and $\Psi^{(k)}(y,\theta_1)$ are
 N=1 chiral superfields.

The soft breaking of N=2 SUSY down to N=0 can be achieved by
shifting the imaginary part of $\tau$:
\begin{equation}
{\mathcal I}m \tau  \to {\mathcal I}m\tilde{\tau} ={\mathcal
I}m\tau (1+ M_1\theta_1^2+
M_2\theta_2^2+M_3\theta_1^2\theta_2^2)\label{shift}
\end{equation}
This leads to
\begin{eqnarray}
\Delta{\mathcal L}&=& \left[ -\frac{M_1}{4}\lambda\lambda
-\frac{M_2}{4}\psi\psi - (\frac{M_1M_2}{4}-\frac{M_3}{4})\phi\phi
+h.c.\right] -(\frac{M_1^2}{4}+\frac{M_2^2}{4})\bar \phi\phi
\nonumber,
\end{eqnarray}
where the fields $\lambda$ are the gauginos, $\psi$ and $\phi$ are
the spinor and scalar matter  fields, respectively.

Now one can use the power of duality in N=2 SUSY theory and take
the
 Seiberg-Witten solution~\cite{SW}
\begin{equation}
\tau = \frac{da_D}{du}/\frac{da}{du}, \label{SW}
\end{equation}
where
\begin{eqnarray*}
a_D(u)&=&\frac{i}{2}(u-1)F(\frac 12,\frac 12,2;\frac{1-u}{2}),\
a(u)=\sqrt{2(1+u)}F(-\frac 12,\frac 12,1;\frac{2}{1+u}).
\end{eqnarray*}

In perturbative domain when $u \sim Q^2/\Lambda^2 \to \infty, \
a=\sqrt{2u}, \ \ a_D=\frac{i}{\pi}a(2\ln a +1)$ one reproduces the
well known one-loop result
\begin{equation}
\frac{4\pi}{g^2}=\frac{1}{\pi}[\ln \frac{Q^2}{\Lambda^2}+3].
\end{equation}
 Assuming that
renormalizations in N=2 SUSY theory follow the properties of those
in N=1, one can try to apply the same expansion procedure for a
non-perturbative solution. Substituting eq.(\ref{shift})
 into (\ref{SW}) with
 $$u\to \tilde{u} =u(1+M_1^0\theta_1^2+ M_2^0\theta_2^2+M_3^0\theta_1^2\theta_2^2)$$
  and expanding over
$\theta_1^2$ and $\theta_2^2$,
 one gets an analog of S-W solution for the mass terms~\cite{Bog}:
$$ M_1=M_1^0\frac{\displaystyle {\mathcal I}m
\left[u\left(\frac{a_D''}{a_D'} -\frac{a''}{a'}\right)\tau
\right]}{\displaystyle  {\mathcal I}m \ \tau}\, ,\ \ \ \ \
M_2=M_2^0\frac{\displaystyle {\mathcal I}m
\left[u\left(\frac{a_D''}{a_D'} -\frac{a''}{a'}\right)\tau
\right]}{\displaystyle  {\mathcal I}m \ \tau}\, , $$
$$ M_3=\frac{\displaystyle {\mathcal I}m
\left[M_3^0u\!\left(\frac{a_D''}{a_D'}\!-\!
\frac{a''}{a'}\right)\tau\!
+\!M_1^0M_2^0u^2\!\left(\frac{a_D'''}{a_D'}\!-\!\frac{a'''}{a'}\!+2
\!\frac{a_D''}{a_D'}\left(\frac{a_D''}{a_D'}-
\frac{a''}{a'}\right)\right)\tau \right]}{\displaystyle {\mathcal
I}m \ \tau} $$
In perturbative regime one has
\begin{eqnarray*}
 M_1&=&\frac{M_1^0}{\ln Q^2/\Lambda^2+3}, \ \ \ \
 M_2=\frac{M_2^0}{\ln Q^2/\Lambda^2+3}, \ \ \ \
  M_3=\frac{M_3^0-M_1^0M_2^0}{\ln Q^2/\Lambda^2+3}.
\end{eqnarray*}

\section{Renormalization of the Fayet-Iliopoulos Term}

We gave above a complete set of the rules needed for writing down
the RG equations for the soft SUSY breaking terms in an arbitrary
non-Abelian N=1 SUSY gauge theory. However, in the Abelian case,
there exists an additional gauge invariant term, the so-called
Fayet-Iliopoulos or the D-term~\cite{FI}
\begin{equation}
{\mathcal L}_{F.I.}=\xi D=\int d^4\theta \xi V,  \label{FI}
\end{equation}
which requires special consideration. In Ref.~\cite{FNPRS}, it has
been shown that in the unbroken theory this term is not
renormalized provided the sum of hypercharges and their cubes
equals zero. These requirements guarantee the absence of chiral
and gravity anomalies and are usually satisfied in realistic
models.

In case of a softly broken Abelian SUSY gauge theory, the F-I term
happens to be renormalized even if anomalies are cancelled. The RG
equation for $\xi$ depends not only on itself, but on the other
soft breaking parameters (the soft mass of chiral scalars $m^2$,
the soft triple coupling $A^{ijk}$ and the gaugino masses $M_i$).
Recently, the renormalization  of $\xi$  has been performed up to
three loops~\cite{JJPar} using the component approach and/or
superfields with softly broken Feynman rules. Following our main
idea that renormalizations of a softly broken SUSY theory are
completely defined by a rigid one, we argue that the
renormalization of the F-I term, in full analogy with all the
other soft terms renormalizations, is completely defined in a
rigid or an unbroken theory. However, contrary to the other soft
renormalizations, there is no simple differential operator that
acts on the renormalization functions of a rigid theory and allows
one to get the renormalization of the F-I term. One needs an
analysis of the superfield diagrams and some additional diagram
calculations in components.

The addition of the F-I term leads to the modification of the
Lagrangian in components. The relevant part of the Lagrangian is
\begin{equation}
{\mathcal L} =\displaystyle{ \frac{1}{2\, g^2}}D^2 +  \xi D +
D\bar \phi^{\,j} {\mathcal Y}_j^i\phi_i - \bar \phi^{\,j}
(m^2)_j^i\phi_i+...
\end{equation}
where  ${\mathcal Y\,}^i_j$ is the hypercharge matrix of the
chiral supermultiplet, and $(m^2)^i_j$ is a soft scalar mass.
After eliminating the auxiliary field $D$ this becomes
\begin{equation}
{\mathcal L} = -\bar\phi^{\,j} (m^2)_j^i\phi_i -
\displaystyle{\frac{1}{2}}g^2(\bar \phi^{\,j}{\mathcal
Y}_j^i\phi_i)^2+..., \label{ldel}
\end{equation}
where
\begin{equation}
 (m^2)^i_j = (m^2)^i_j + g^2 \xi {\mathcal
Y}^i_j. \label{mbar}
\end{equation}
From eqs.(\ref{ldel}) and (\ref{mbar}) it follows that the F-I
term gives an additional contribution to the renormalization of
the soft scalar mass $(m^2)^i_j$
\begin{equation}
[\beta_{m^2}]^i_j = [\beta_{m^2}]^i_j + \beta_{g^2} \xi {\mathcal
Y}^i_j+g^2 \beta_\xi(m^2,...) {\mathcal
Y}^i_j=[\beta_{m^2}]^i_j+g^2 \beta_\xi(m^2,...) {\mathcal Y}^i_j.
\end{equation}
The last equality follows from the fact that eq.(\ref{ldel}) does
not contain $\xi$ explicitly and, hence, $\xi$ should be dropped
from all the expressions.

 There are four different types of contributions to
the renormalization of the F-I term in a softly broken theory:
those proportional to $(m^2)^i_j$, $M \bar M$, $A^{ijk} {\bar
A}_{lmn}$ and $M {\bar A}_{lmn}$ ($\bar M A^{ijk}$).

We have found that all the information about the renormalizations
of the soft SUSY breaking terms  is contained in a rigid, unbroken
theory. To calculate the renormalization of an additional
Fayet-Iliopoulos term, one needs an analysis of superfield
diagrams. To find the contribution proportional to the soft scalar
mass $(m^2)^i_j$ (the square of gaugino mass $M \bar M$), one
needs to take the self-energy diagrams for the vector superfield
and replace one of the external vertices with the hypercharge
${\mathcal Y}^i_j$ by $(m^2)^i_j$ ($M \bar M \delta^i_j$). In this
case, there is no need to do any calculations except in
superfields.

The other contributions (proportional to $A\bar A$ and $M \bar A$)
can be found from the analysis of the matter superfield propagator
diagrams in a rigid theory and the corresponding component
diagrams in a softly broken theory extracting from the latter  the
contribution of the tadpole graphs. In this case, one needs to
calculate additionally some component diagrams the number of which
is essentially reduced compared to a direct component
calculation~\cite{KV2}.

\section{Conclusion}

Summarizing, we would like to stress once again that is very
useful to consider  a spontaneously broken theory  in terms of a
rigid one in an external field. In case when one is able to absorb
the external field into the redefinition of parameters of the
original theory and perform the renormalizations for an arbitrary
field, one can reproduce renormalization properties of a
spontaneously broken theory from a rigid one. The Grassmannian
expansion in softly broken SUSY theories happens to be a very
efficient and powerful  method which can be applied in various
cases where the renormalization procedure in concerned. It
demonstrates once again that softly broken SUSY theories  are
contained in rigid  ones and inherit their renormalization
properties.

The following statements are valid:
\begin{itemize}\item
 All the renormalizations are defined in a rigid
theory.  There are no  independent  renormalizations in a softly
broken theory.
\item
RG flow in a softly broken theory follows that in a rigid theory.
\item This statement is true for RG equations, solutions to these equations,
particular (fixed point) solutions, approximate solutions, etc.
\item Renormalization of the F-I term needs a special treatment
but can be also deduced from unbroken theory.
\end{itemize}

\section*{Appendix A. Three-loop renormalizations in the MSSM}
\setcounter{equation}0
\renewcommand{\theequation}{A.\arabic{equation}}

In this section, we present explicit formulae for rigid and soft
term renormalizations in the MSSM in  the three-loop approximation
in the case when we retain only $\alpha_3$ and top Yukawa coupling
$Y_t$.

The rigid renormalizations are~\cite{Ferreira}
\begin{eqnarray}
\beta_{\alpha_3}&=&-3\alpha_3^2+\alpha_3^2(14\alpha_3-4Y_t)+
\alpha_3^2[\frac{347}{3}\alpha_3^2-\frac{104}{3}\alpha_3Y_t+30Y_t^2], \\
\gamma_t&=&(2Y_t-\frac{8}{3}\alpha_3)-(8Y_t^2+\frac{8}{9}\alpha_3^2)
+[(30+12\zeta_3)Y_t^3 \\ &+&
(\frac{16}{3}+96\zeta_3)Y_t^2\alpha_3-
(\frac{64}{3}+\frac{544}{3}\zeta_3)Y_t\alpha_3^2+(\frac{2720}{27}+320
\zeta_3)\alpha_3^3], \nonumber\\
\gamma_b&=&-\frac{8}{3}\alpha_3-\frac{8}{9}\alpha_3^2 +
[-\frac{80}{3}Y_t\alpha_3^2+(\frac{2720}{27}+320
\zeta_3)\alpha_3^3], \\
\gamma_Q&=&(Y_t-\frac{8}{3}\alpha_3)-(5Y_t^2+\frac{8}{9}\alpha_3^2)+
[(15+6\zeta_3)Y_t^3 \\
&+&(\frac{40}{3}+48\zeta_3)Y_t^2\alpha_3-
(\frac{72}{3}+\frac{272}{3}\zeta_3)Y_t\alpha_3^2+(\frac{2720}{27}+320
\zeta_3)\alpha_3^3], \nonumber\\
\gamma_{H_2}&=&(3Y_t)-(9Y_t^2-16Y_t\alpha_3) \\ &+&
[(57+18\zeta_3)Y_t^3+(72-144\zeta_3)Y_t^2\alpha_3-
(\frac{160}{3}+16\zeta_3)Y_t\alpha_3^2], \nonumber\\
\beta_{Y_t}&=&Y_t\left\{(6Y_t-\frac{16}{3}\alpha_3)-(22Y_t^2-16Y_t\alpha_3
+\frac{16}{9}\alpha_3^2)\right. + [(102+36\zeta_3)Y_t^3
\nonumber\\ &+& \left. \frac{272}{3}Y_t^2\alpha_3-
(\frac{296}{3}+288\zeta_3)Y_t\alpha_3^2+(\frac{5440}{27}+640
\zeta_3)\alpha_3^3]\right\}, \\
\beta_{\mu^2}&=&\mu^2\left\{3Y_t-(9Y_t^2-16Y_t\alpha_3)\right.  \\
&+& \left. [(57+18\zeta_3)Y_t^3+(72-144\zeta_3)Y_t^2\alpha_3-
(\frac{160}{3}+16\zeta_3)Y_t\alpha_3^2]\right\} \nonumber,
\end{eqnarray}

Using the explicit form of  anomalous dimensions calculated up to
some order, one can reproduce the desired RG equations for the
soft terms. In case of squark and slepton masses, they contain the
contributions from unphysical masses $\Sigma_{\alpha_i}$. To
eliminate them, one has to  solve the equation for
$\Sigma_{\alpha_i}$. In the case of the MSSM up to three loops,
the solutions are~\cite{KV}
\begin{eqnarray}
  {\Sigma}_{\alpha_1}&=&M_1^2-\alpha_1 \sigma_1 -\frac{199}{25}{\alpha}_1^2
M_1^2-\frac{27}{5}\alpha_1\alpha_2M_2^2
-\frac{88}{5}\alpha_1\alpha_3M_3^2\nonumber \\
&+&\frac{13}{5}\alpha_1Y_t(\Sigma_t+A_t^2)+\frac{7}{5}\alpha_1Y_b(\Sigma_b+A_b^2)
+\frac{9}{5}\alpha_1Y_{\tau}(\Sigma_{\tau}+A_{\tau}^2),\\
  {\Sigma}_{\alpha_2}&=&M_2^2-\alpha_2 (\sigma_2-4M_2^2) -{\alpha}_2^2(
4\sigma_2+9M_2^2)-\frac{9}{5}\alpha_2\alpha_1M_1^2
-24\alpha_2\alpha_3M_3^2\nonumber \\
&+&3\alpha_2Y_t(\Sigma_t+A_t^2)+3\alpha_2Y_b(\Sigma_b+A_b^2)
+\alpha_2Y_{\tau}(\Sigma_{\tau}+A_{\tau}^2),\\
{\Sigma}_{\alpha_3}&=&M_3^2-\alpha_3 (\sigma_3-6M_3^2)
-{\alpha}_3^2( 6\sigma_3-22M_3^2)
-\frac{11}{5}\alpha_3\alpha_1M_1^2-9\alpha_3\alpha_2M_2^2\nonumber \\
&+&2\alpha_3Y_t(\Sigma_t+A_t^2)+2\alpha_3Y_b(\Sigma_b+A_b^2),
\end{eqnarray}
where we have used the combinations~\cite{MV}
\begin{eqnarray}
\sigma_1 &=&\frac 15 \left[3 ( m^2_{H_1} +   m^2_{H_2})+ 3(
 \tilde m^2_{Q}+3 \tilde m^2_{L} +8 \tilde m^2_{U}+2 \tilde m^2_{D}+6
 \tilde m^2_{E})\right],\nonumber\\
\sigma_2 &=& m^2_{H_1} +   m^2_{H_2}+ 3(
 3\tilde m^2_{Q}+ \tilde m^2_{L}),\label{s2}\\
\sigma_3 &=& 3(
 2\tilde m^2_{Q}+ \tilde m^2_{U}+ \tilde m^2_{D}),\nonumber\\
 \Sigma_t&=&\tilde{m}^2_t+\tilde{m}^2_Q+m^2_{H_2},\
 \Sigma_b =\tilde{m}^2_b+\tilde{m}^2_Q+m^2_{H_1}, \
 \Sigma_\tau =\tilde{m}^2_\tau+\tilde{m}^2_L+m^2_{H_1}.
 \nonumber
\end{eqnarray}

The corresponding soft term renormalizations read
\begin{eqnarray*}
\beta_{M_3}&=&-3\alpha_3M_3+28\alpha_3^2M_3-4Y_t\alpha_3(M_3-A_t)\\
&+& 347\alpha_3^3M_3-\frac{104}{3}\alpha_3^2Y_t(2M_3-A_t)
+30\alpha_3Y_t^2(M_3-2A_t),\nonumber\\
\beta_{A_t}&=&(6Y_tA_t+\frac{16}{3}\alpha_3M_3)-[44Y_t^2A_t-16Y_t\alpha_3
(A_t-M_3) -\frac{32}{9}\alpha_3^2M_3]\nonumber  \\ &+&
[(306+108\zeta_3)Y_t^3A_t+\frac{272}{3}Y_t^2\alpha_3(2A_t-M_3)\\
&-&(\frac{296}{3}+288\zeta_3)Y_t\alpha_3^2(A_t-2M_3)
-3(\frac{5440}{27}+640 \zeta_3)\alpha_3^3M_3],\nonumber \\
 \beta_{B}&=&3Y_tA_t-[18Y_t^2A_t-16Y_t\alpha_3(A_t-M_3)]
+[(171+54\zeta_3)Y_t^3A_t
\\&+&(72-144\zeta_3)Y_t^2\alpha_3(2A_t-M_3)
 -(\frac{160}{3}+16\zeta_3)Y_t\alpha_3^2(A_t-2M_3)],\nonumber \\
\beta_{\tilde{m}^2_t}&=&2Y_t(\Sigma_t+A_t^2)-\frac{16}{3}\alpha_3M_3^2
-16Y_t^2(\Sigma_t+2A_t^2)-\frac{64}{3}\alpha_3^2M_3^2+
 \frac 83 \alpha_3^2\sigma_3\nonumber \\
&+& 3(30+12\zeta_3)Y_t^3(\Sigma_t+3A_t^2)
+(\frac{16}{3}+96\zeta_3)Y_t^2\alpha_3 [(2A_t-M_3)^2 \\
&+&2\Sigma_t+M_3^2]-
(\frac{64}{3}+\frac{544}{3}\zeta_3)Y_t\alpha_3^2
[(A_t-2M_3)^2+\Sigma_t+2M_3^2]  \nonumber \\
&-&\frac{16}{3}Y_t\alpha_3^2(\Sigma_t+A_t^2)+
4(\frac{2564}{9}+960\zeta_3)\alpha_3^3M_3^2+\frac{160}{9}\alpha_3^3\sigma_3, \nonumber\\
\beta_{\tilde{m}^2_b}&=&-\frac{16}{3}\alpha_3M_3^2-\frac{64}{3}\alpha_3^2M_3^2+
\frac 83 \alpha_3^2\sigma_3  - \frac{80}{3}Y_t\alpha_3^2
[(A_t-2M_3)^2+\Sigma_t\nonumber \\&+&2M_3^2]
-\frac{16}{3}Y_t\alpha_3^2(\Sigma_t+A_t^2)+
4(\frac{2564}{9}+960\zeta_3)\alpha_3^3M_3^2+\frac{160}{9}\alpha_3^3\sigma_3, \\
\beta_{\tilde{m}^2_Q}&=&Y_t(\Sigma_t+A_t^2)-\frac{16}{3}\alpha_3M_3^2
-10Y_t^2(\Sigma_t+2A_t^2)- \frac{64}{3}\alpha_3^2M_3^2+ \frac 83
\alpha_3^2\sigma_3\nonumber \\&+&
3(15+6\zeta_3)Y_t^3(\Sigma_t+3A_t^2)
+(\frac{40}{3}+48\zeta_3)Y_t^2\alpha_3 [(2A_t-M_3)^2\\ &+&
2\Sigma_t+M_3^2] -
(\frac{72}{3}+\frac{272}{3}\zeta_3)Y_t\alpha_3^2
[(A_t-2M_3)^2+\Sigma_t+2M_3^2] \nonumber \\
&-&\frac{16}{3}Y_t\alpha_3^2(\Sigma_t+A_t^2)+
4(\frac{2564}{9}+960\zeta_3)\alpha_3^3M_3^2+\frac{160}{9}\alpha_3^3\sigma_3,\nonumber \\
\beta_{m^2_{H_2}}&=&3Y_t(\Sigma_t+A_t^2) -18Y_t^2(\Sigma_t+2A_t^2)
+ 16Y_t\alpha_3 [(A_t-M_3)^2+\Sigma_t+M_3^2]\nonumber
\\ &+&
3(57+18\zeta_3)Y_t^3(\Sigma_t+3A_t^2)
+(72-144\zeta_3)Y_t^2\alpha_3 [(2A_t-M_3)^2
\\&+&2\Sigma_t+M_3^2]
 - (\frac{160}{3}+16\zeta_3)Y_t\alpha_3^2
[(A_t-2M_3)^2+\Sigma_t+ 2M_3^2]\nonumber \\
&-&16Y_t\alpha_3^2(\sigma_3-6M_3^2).\nonumber
\end{eqnarray*}

\section*{Appendix B}
\setcounter{equation}0
\renewcommand{\theequation}{B.\arabic{equation}}

The RG equation for the parameter $\xi$ in a rigid theory is
\begin{equation}\label{rxi}
  \dot \xi = - \gamma_V \xi,
\end{equation}
where $\gamma_V$ is the anomalous dimension of the gauge
superfield. To find the soft terms $x,\bar x$ and $z$, one should
solve the modified equation
\begin{equation}\label{rxim}
  \dot {\tilde \xi} = - \gamma_V(\tilde \alpha, \tilde y, \tilde \xi) \tilde
  \xi.
\end{equation}

In one-loop order  $\gamma_V= (b_1+b_2\xi) \alpha$, where
$b_1+b_2=Q$,  and the solutions are
\begin{eqnarray}
   x&=& -(M+x_0)\frac{b_1+b_2\xi}{Q}\ , \ \ \ \ \bar x =  -(\bar M +\bar
   x_0)\frac{b_1+b_2\xi}{Q}\ , \nonumber \\
    z &=& -(\Sigma_\alpha+z_0)\frac{b_1+b_2\xi}{Q} + \frac{b_2\xi}{Q}(M+x_0)(\bar M
    +\bar x_0)\frac{b_1+b_2\xi}{Q}\ ,
\end{eqnarray}
where  $x_0, \bar x_0$, and $z_0$ are arbitrary constants. In the
Abelian case when $b_1=Q, \ b_2=0$, the solutions are simplified
and  can be chosen as
$$x=-M(1-\xi),\ \ \ \bar x = -\bar M (1-\xi), \ \ \  z=-\Sigma_\alpha (1-\xi)-
M\bar M \xi (1-\xi).$$ Together with the expression for $\tilde
\alpha$ (\ref{Tildeg}) it gives eq.(\ref{ghs2}) above.

\section*{Acknowledgments}

I would like to thank my colleagues L.Avdeev, I.Kondrashuk,
S.Codoban, G.Moultaka and V.Velizhanin in collaboration with whom
these results have been obtained. I am grateful to the organizers
of the conference "Continuous Advances in QCD-02" and especially
to M.Shifman for their invitation to participate in the conference
and to present this report. Many thanks are to M.Shifman,
A.Vainshtein, M.Strassler, I.Jack, T.Jones and R.Rattazzi for
useful discussions.

Financial support from RFBR grants \# 02-02-16889 and \#
00-15-96691 is kindly acknowledged.


\end{document}